\documentclass[aps,nofootinbib,prl,superscriptaddress,tightenlines,notitlepage,
twocolumn,showpacs,floatfix]{revtex4-1}
\usepackage[colorlinks]{hyperref}
\usepackage{tabularx}
\usepackage{bm}
\usepackage{graphicx}
\usepackage{amssymb,bm,tensor}
\usepackage{xcolor}
\usepackage{amsmath}
\usepackage[varg]{txfonts}
\usepackage{enumerate}
\usepackage[normalem]{ulem}
\usepackage{bm}
\usepackage[OT1]{fontenc}

\newcommand{\reply}[1]{ #1}

\begin{document}

\title{Measuring the Spin of the Galactic Center Supermassive Black Hole with Two Pulsars}

\date{\today}

\author{Zexin Hu}
\affiliation{Department of Astronomy, School of Physics, Peking University, 
Beijing 100871, China}
\affiliation{Kavli Institute for Astronomy and Astrophysics, Peking University, 
Beijing 100871, China}

\author{Lijing Shao}\email{lshao@pku.edu.cn}
\affiliation{Kavli Institute for Astronomy and Astrophysics, Peking University, 
Beijing 100871, China}
\affiliation{National Astronomical Observatories, Chinese Academy of Sciences, 
Beijing 100012, China}

\begin{abstract}
As a key science project of the Square Kilometre Array (SKA), the discovery and
timing observations of radio pulsars in the Galactic Center would provide
high-precision measurements of the spacetime around the supermassive black hole,
Sagittarius~A* (Sgr~A*), and initiate novel tests of general relativity. The
spin of Sgr A* could be measured with a relative error of $\lesssim 1\%$ by
timing one pulsar with timing precision that is achievable for the SKA. However,
the real measurements depend on the discovery of a pulsar in a very compact
orbit, $P_b\lesssim0.5\,{\rm yr}$. Here for the first time we propose and
investigate the possibility of probing the spin of Sgr~A* with two or more
pulsars that are in orbits with larger orbital periods, $P_b\sim 2- 5\,{\rm
yr}$, which represents a more realistic situation from population estimates. We
develop a novel method for directly determining the spin of Sgr~A* from the
timing observables of two pulsars and it can be readily extended for combining
more pulsars. With extensive mock data simulations, we show that combining a
second pulsar improves the spin measurement by $2-3$ orders of magnitude in some
situations, which is comparable to timing a pulsar in a very tight orbit.
\end{abstract}

\maketitle

\textit{Introduction.}---Black holes (BHs) are the most intriguing objects for
testing general relativity (GR). While the detection of gravitational waves from
stellar-mass BH mergers~\cite{KAGRA:2021vkt} provided abundant knowledge of the
highly dynamical, strong-field regime of gravity, the supermassive BHs (SMBHs)
existing at the center of most massive galaxies~\cite{McConnell:2012hz,
Kormendy:2013dxa} represent another ideal laboratory for BH physics.  The SMBH
dwelling in our Galactic Center (GC)~\cite{Ghez:2008ms, Genzel:2010zy,
EventHorizonTelescope:2022wkp}, Sagittarius~A* (Sgr~A*), with the largest mass
to distance ratio among the known BHs, is the most promising target for
precision observations and unique tests of strong-field
gravity~\cite{EventHorizonTelescope:2022xqj}.

Previous studies have shown that timing observations of radio pulsars in compact
orbits around Sgr~A* would allow us to probe the spacetime around this SMBH to 
unprecedented accuracy and provide us with unique tests of the no-hair theorem 
for Kerr BHs~\cite{Wex:1998wt, Liu:2011ae,Psaltis:2015uza,
Zhang:2017qbb,Bower:2018mta,Dong:2022zvh,Hu:2023ubk,Hu:2023vsg}. Regular timing
of a normal pulsar in a highly eccentric orbit ($e\gtrsim 0.8$) tightly around
Sgr~A* ($P_b\lesssim 0.5\,{\rm yr}$) will measure the spin and quadrupole moment
of Sgr~A* to a relative precision of about $1\%$ within five years for a timing
precision of $1\,{\rm ms}$, which is achievable for future radio
telescopes~\cite{Liu:2011ae, Shao:2014wja, Psaltis:2015uza, Hu:2023ubk}.
However, the existence and detectability of such pulsars in very close orbits
around Sgr~A* is still unclear. The large dispersion measures and the scattering
caused by highly turbulent interstellar medium in the GC make observations favor
high radio frequencies~\cite{Cordes:2002wz}, which then render pulsars
flux-limited because of the steep spectra of emission. Nevertheless, theoretical
models suggest that there can be $10^2-10^3$ pulsars orbiting around Sgr~A* with
$P_b\lesssim 100\,{\rm yr}$~\cite{Pfahl:2003tf, Zhang:2014kva,2020A&A...641A.102S}, and present
observations including the discoveries of a magnetar~\cite{Eatough:2013nva,
Eatough:2022xmg}, a millisecond pulsar~\cite{Lower:2024sdi} and several normal
pulsars~\cite{Johnston:2006fx,Deneva:2009mx} in that region all suggest that a
large population of pulsars are awaiting for discoveries in the GC for future
high-frequency surveys.

Measuring the BH spin is crucial for gravity tests.  \reply{Determining the spin of
Sgr~A* based on the timing of a single pulsar suffers from the so-called leading-order degenracy among spin parameters~\cite{Liu:2011ae,Zhang:2017qbb}.}  The measurable secular
changes of the pulsar orbit in a short time, caused by the frame-dragging
effect~\cite{Barker:1975ae}, are the linear rate of the periastron precession,
$\dot{\omega}$, and the linear change of the projected semimajor axis of the
pulsar orbit, $\dot{x}$. However, these two quantities cannot fully determine
the BH spin which has three components. To break the degeneracy, one needs at
least one accurate measurement of  higher-order time derivatives such as 
$\ddot{\omega}$ or $\ddot{x}$~\cite{Liu:2011ae}, which in general requires a
much longer observation time span or a pulsar in an orbit with very short 
orbital period whose relativistic effects are large. 

There are still two possible ways to break the leading-order degeneracy and 
provide a better spin determination without requiring a pulsar in a rare orbital
situation. One is to combine the measurement of the proper motion of the pulsar
from astrometric observations~\cite{Zhang:2017qbb}.  \reply{One can infer the secular
change in the longitude of the ascending node, $\dot{\Omega}$, from the proper motion of the pulsar and
provide an additional constraint on the spin of the central BH.} However, an
expected astrometric accuracy in the order of $10\, \mu{\rm
as}$~\cite{Fomalont:2004hr} is still not enough and contributes little when 
combining with the timing data~\cite{Zhang:2017qbb}. The other possibility is to
combine the observation of another pulsar with a different orbital inclination
$i$, as the leading-order degeneracy mainly depends on $i$~\cite{Zhang:2017qbb}.
Considering a potentially large pulsar population in the GC, finding two or more
pulsars orbiting around Sgr~A* in the future may provide us a more likely way in
measuring the spin of Sgr~A*.

In this {\it Letter}, we develop a novel method for determining the spin of
Sgr~A* by combining  observations of two or more pulsars. We demonstrate that
for a pulsar with orbital period $P_b\sim 2-5\,{\rm yr}$, combining a second
pulsar can improve the spin measurement significantly. While previous studies
focused on timing one pulsar in a very tight orbit ($P_b\lesssim 0.5\,{\rm
yr}$), we show that finding two or more pulsars with larger orbital periods are
more likely to be the case for future observations.

\textit{Combining two pulsars.}---When the changes in the Keplerian parameters
of a pulsar orbit are slow over the observation time span, the direct way to
characterize  secular effects is to fit the timing data with timing parameters
varying linearly in time, such as  $\omega=\omega_0+\dot{\omega}\,t$ and 
$x=x_0+\dot{x}\,t$~\cite{Damour:1991rd}. Different from the binary pulsar systems that are regularly timed at present, in which even for the very precisely timed pulsars only some of them have measurents of $\dot{\omega}$~\cite{Chakrabarti:2020abx,Donlon:2024ugk}, in a pulsar-SMBH system, the leading-order drivatives will be measured soon after the start of observation
and have a rather high measurement precision due to the large relativistic effects from the SMBH~\cite{Liu:2011ae,Psaltis:2015uza,Zhang:2017qbb,Hu:2023ubk}. Here we use the two combinations
introduced by~\citet{Liu:2011ae} for the leading-order derivatives and other
timing observables, which are $\mathcal{X}\equiv -\dot{x}s_i^2
\big(x\hat{\Omega} \big)^{-1}$ and $\mathcal{W}\equiv(\dot{\omega}-
\dot{\omega}_M)s_i^2\hat{\Omega}^{-1}$, where $s_i=\sin i$,
$\hat{\Omega}=4\pi\beta_O^3/P_b(1-e^2)^{3/2}$ and 
$\dot{\omega}_M=6\pi\beta_O^2/P_b(1-e^2)$, with $\beta_O=\big(2\pi GM/c^3P_b
\big)^{1/3}$ the orbital velocity parameter introduced by~\citet{Damour:1991rd}.
Note that $s_i$ and $M$ in these combinations can be measured precisely from the
Shapiro delay~\cite{Shapiro:1964uw} that is large enough here even for face-on
orbits~\cite{ Liu:2011ae, Hu:2023ubk}. 

To combine the timing results of two pulsars and determine the spin of the 
central BH, one has to translate the measurement constraints of both pulsars in
a common parameter space. In this case, the natural choice is the
$\chi$-$\chi_\lambda$ plane, where $\chi_\lambda$ is the projection of the
dimensionless spin $\chi$ of the BH in the line of sight direction. \reply{After some tedious simplification, we show that, for each
pulsar $k$ ($k=1,2$), timing observations give one constraint represented by a hyperbola,
\begin{equation}\label{eq:chi}
    \frac{\mathcal{X}_k^2}{s_{i_k}^2(1-s_{i_k}^2)}+\frac{\mathcal{W}_k^2}{s_{i_k}^2(1+3s_{i_k}^2)}=
    \chi^2-\frac{1+3s_{i_k}^2}{(1-3s_{i_k}^2)^2}\left(\chi_k+\frac{3
    \mathcal{W}_kc_{i_k}}{1+3s_{i_k}^2}\right)^2\,,
\end{equation}
where $c_{i_k}=\cos{i_k}$.}  If one  observes two or more pulsars, all these curves in
the $\chi$-$\chi_\lambda$ plane should intersect at one point that gives $\chi$
and $\chi_\lambda$. 

\begin{figure}
    \centering
    \includegraphics[width=8cm]{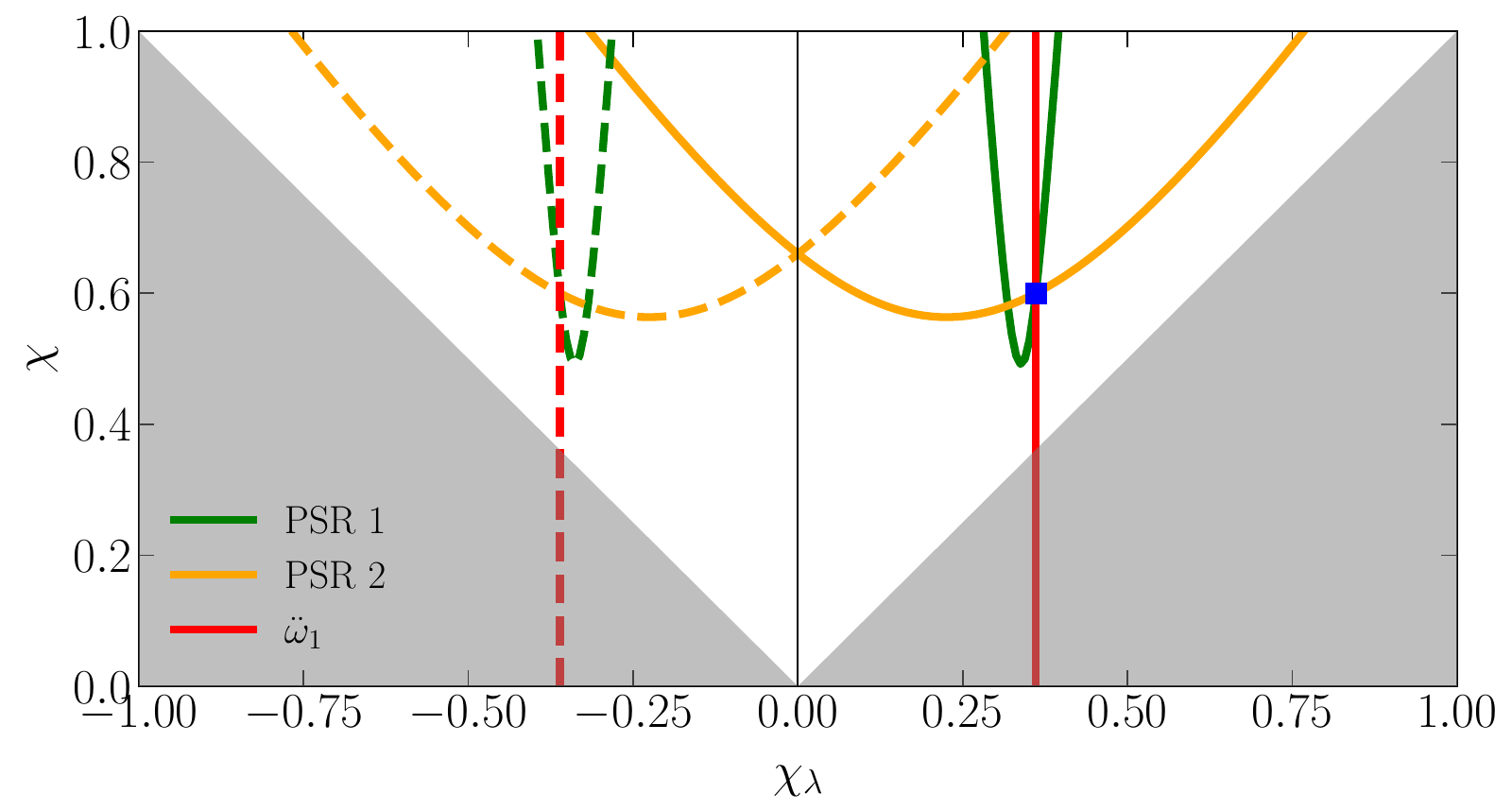}
    \caption{Illustration of the determination of the spin of Sgr~A* with  
    measurements of two pulsars in the $\chi$-$\chi_\lambda$ plane. \reply{The square symbol located at $(0.36,0.6)$ marks the assumed true value of the BH spin.} The solid curves show
    the constraints from two pulsars, while the dashed lines correspond to the
    $i\leftrightarrow \pi-i$ ambiguity. Shaded region is excluded by the
    condition $\left|\chi_\lambda\right|\leq\chi$.}
    \label{fig:ana}
\end{figure}

Figure~\ref{fig:ana} shows an illustration of using the timing of two pulsars to
determine the spin of Sgr~A* in the $\chi$-$\chi_\lambda$ plane. \reply{The system 
parameters are randomly chosen to avoid the accidental degeneracy which will be discussed further in the numerical simulation section.} \reply{For each pulsar $\alpha$, the measurements of $({\cal X}_k, {\cal W}_k)$
give a hyperbola shown by the solid curve, and the $i_k\leftrightarrow \pi-i_k$
ambiguity gives another hyperbola shown by the dashed line.} In general, there
can exist at most eight different solutions for $(\chi, \chi_\lambda)$.  A rough
measurement of any second-order derivatives, for example, the $\ddot{\omega}$ 
for one of the pulsar, as shown by the vertical red line, or timing of a third 
pulsar can reduce this degeneracy into two solutions whose $\chi$ has the same
value and $\chi_\lambda$ has opposite signs.

The methodology described above can be easily extended for combining more
pulsars.  For each additional pulsar, it provides an independent test of GR as
all these curves should intersect at one point within their measurement
uncertainties. On the other hand, if one assumes that GR is correct, timing more
pulsars can be a unique probe for us to explore and model the (stellar and gas)
mass distribution near the GC~\cite{Merritt:2009ex,Liu:2011ae,Hu:2023ubk,Carleo:2023qxu}. As
only the leading-order time derivatives and a rough value of a higher-derivative
parameter are used, one is expected to obtain a high-precision measurement after
a short time observing two pulsars.

\textit{Numerical simulation.}---To further explore the method introduced above
and give a quantitative description of the expected measurement precision in
measuring the spin of Sgr~A* by timing two pulsars,  we perform mock data
simulations based on the numerical timing model for pulsar-SMBH systems
developed earlier~\cite{Hu:2023ubk}. In this timing model, we numerically
integrate the pulsar orbital motion based on the post-Newtonian (PN) equation of
motion, $\ddot{\bm{r}}=-GM\hat{\bm{r}}/r^2+\ddot{\bm{r}}_{\rm 1PN}
+\ddot{\bm{r}}_{\rm SO}+\ddot{\bm{r}}_{\rm Q}$, where $\bm{r}$ is the relative
separation pointing from the SMBH to the pulsar, $\ddot{\bm{r}}_{\rm 1PN}$ is
the first PN correction, $\ddot{\bm{r}}_{\rm SO}$ and $\ddot{\bm{r}}_{\rm Q}$
are the spin and quadrupole contribution from the central SMBH
respectively~\cite{Barker:1975ae}, $M$ is the mass of the SMBH,
$r\equiv|\bm{r}|$, and $\hat{\bm{r}}\equiv\bm{r}/r$. Taking advantage of the PN
expansion, we  treat the dimensionless spin $\chi$ and the dimensionless
quadrupole moment $q$ of the SMBH as independent parameters, which renders a
test of the no-hair theorem $q=-\chi^2$ for Sgr A* possible~\cite{Hu:2023ubk}.
As the mass ratio $m_{\rm PSR}/M\lesssim 10^{-6}$, we treat the pulsar as a test
particle in the BH's spacetime. We take into account the R\"{o}mer delay, the
leading-order Shapiro delay and the Einstein delay in the timing
model~\cite{Shapiro:1964uw, Blandford:1976,Damour:1986} and ignore the effects
caused by the proper motion of Sgr~A*~\cite{Shklovskii:1970,Kopeikin:1996}. More
details can be found in Ref.~\cite{Hu:2023ubk}.

The full parameter set, $\bm{\Theta}=\bm{\Theta}_{\rm BH} \cup \bm{\Theta}_{\rm
PSR}$, in the timing model  for a single pulsar includes the parameters of the
SMBH, $\bm{\Theta}_{\rm BH}=\{M,\chi,q,\lambda,\eta\}$, where $\lambda$ and
$\eta$ describe the direction of the spin, and the parameters of the pulsar,
$\bm{\Theta}_{\rm PSR}=\{P_b,e,f,i,\omega,N_0,\nu,\dot{\nu}\}$, where $f$ is the
initial orbital phase and $\{N_0,\nu,\dot{\nu}\}$ relate the pulsar's rotation
number $N$ and pulsar's proper time $T$ via $N(T)=N_0+\nu T+\dot{\nu}T^2/2$.  \reply{We
have set the longitude of the ascending node of the pulsar orbit, $\Omega$, to
be zero as it is not an observable when ignoring the proper motion~\cite{Kopeikin:1994}.} When combining the
timing of two or more pulsars, one should have one set of $\Theta_{\rm PSR}$ for
each pulsar while $\Theta_{\rm BH}$ is common for different pulsars. In
addition, the difference between the longitudes of the ascending node, $\Delta
\Omega=\Omega_2-\Omega_1$, should also be included in the timing
model~\cite{Ransom:2014xla}. As we can have an overall rotation of the system
around the line of sight direction, we fix $\Omega_1=0$ in the following
discussions and we denote the pulsar with a smaller orbital period as PSR1 or
the inner pulsar.

We use the Fisher matrix method for parameter estimation~\cite{Damour:1986}. 
\reply{The log-likelihood function $\mathcal{L}$ in the covariance matrix $C_{\mu
\nu}= \big(\partial^2\mathcal{L}/ \partial\Theta^\mu\partial\Theta^\nu
\big)^{-1}$ is simply the sum from two pulsars,
$\mathcal{L}=\mathcal{L}_1+\mathcal{L}_2$, and 
\begin{equation}
    \mathcal{L}_k=\frac{1}{2\nu_k^2}\sum_{a=1}^{N^{(k)}_{\rm TOA}}
    \frac{\left[N^{(k)}_a(\bm{\Theta})-N^{(k)}_a(\bar{\bm{\Theta}})\right]^2}
    {\left(\sigma^{(k)}_{\rm TOA}\right)^2}\,,\ \ \ k=1,2\,
\end{equation}
where $N^{(k)}_a(\Theta)=N \Big(\Theta;t_a^{(k)\,{\rm TOA}} \Big)$ is the pulsar
rotation number calculated by the timing model corresponding to the $a$-th pulse
arrived at $t_a^{(k)\,{\rm TOA}}$. $\bar{\bm{\Theta}}$ denotes the true system
parameters and $\sigma_{\rm TOA}^{(k)}$ is the timing precision for pulsar $k$.}
As we do not consider the interaction between the two pulsars, the properties of
the Fisher matrix method and the timing model allow one to construct the full
Fisher matrix $F\equiv C^{-1}$ though the combination of the Fisher matrices
$F_i$ of each single pulsar. This simplification is important for combining more
pulsars.

\begin{figure}
    \centering
    \includegraphics[width=8cm]{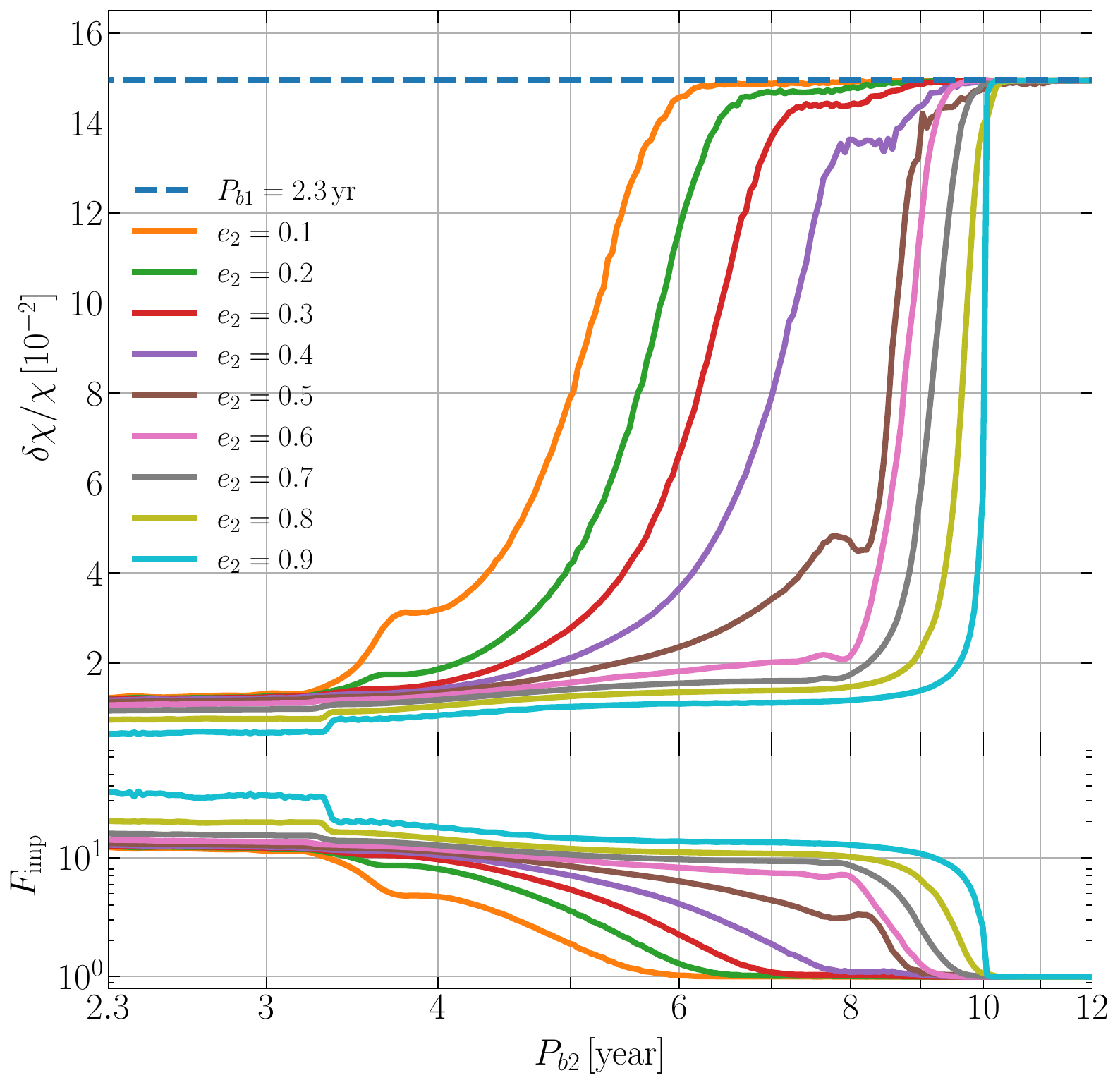}
    \caption{Measurement precision ({\it upper}) and improvement factor ({\it
    lower}) for the spin of Sgr~A* as functions of the second pulsar's orbital
    period ($P_{b2}$) for different orbital eccentricities ($e_2$).  The solid
    curves show the measurement precision from combining the timing of two
    pulsars, while we also show the measurement precision only from the timing
    of the inner pulsar for comparison (dashed line). The initial orbital
    position of the second pulsar is set to be at the apocenter, thus one can
    see at least one pericenter passage for  $P_{b2}<10\,{\rm yr}$.}
    \label{fig:pb}
\end{figure}

We assume a timing precision $\sigma_{\rm TOA}=1\,{\rm ms}$ and an observation 
time span $T_{\rm obs}=5\,{\rm yr}$ for both pulsars, which are realistic for 
future instruments such as the Square Kilometre Array (SKA) and the 
next-generation Very Large Array (ngVLA)~\cite{Weltman:2018zrl, Bower:2018mta}.
In the upper panel of Fig.~\ref{fig:pb} we show the measurement precision of the
spin of Sgr~A* from combining the timing of two pulsars (solid curves) compared 
to the results obtained from the inner pulsar only (dashed line). Here we fix 
the orbital period of the inner pulsar to be $P_{b1}=2.3\,{\rm yr}$ and orbital 
eccentricity $e_1=0.8$ while varying the orbital period $P_{b2}$ and 
eccentricity $e_2$ of the second pulsar. The lower panel of this figure shows 
the improvement factor $F_{\rm imp}=\delta\chi_{\rm single} / \delta\chi_{\rm
combine}$. For this case, the inner pulsar with an orbital period $P_{b1}=2.3
\,{\rm yr}$ can only constrain $\delta \chi / \chi$ to $\sim 15\%$. Combining a 
second pulsar even with a large orbital period can provide an improvement factor
of $\sim 10$ because of the breaking of the leading-order degeneracy.  The
improvement factor from combining a second pulsar will be smaller if the inner
pulsar has a tighter orbit for which itself can break the leading-order
degeneracy. Another case study of $P_{b1}=0.5\,{\rm yr}$ shows an improvement
factor of $2\sim 3$ when combined with a second pulsar whose orbital period
$P_{b2}\lesssim 3\,{\rm yr}$.

\begin{figure}
    \centering
    \includegraphics[width=8cm]{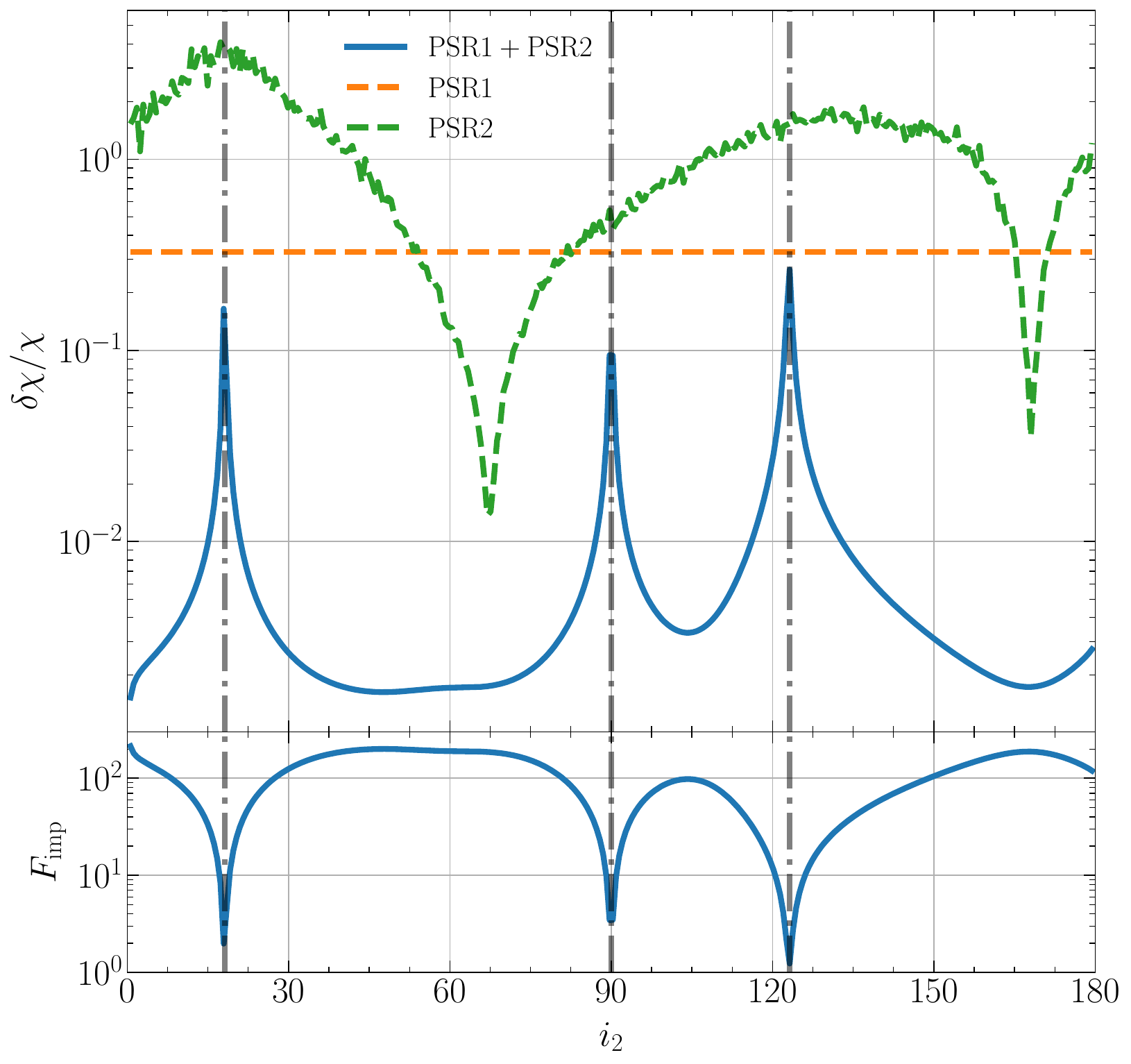}
    \caption{Measurement precision ({\it upper}) of the spin parameter as
    functions of the orbital inclination angle $i_2$ of the second pulsar. Two
    dashed lines show the measurement precision from only one pulsar. The blue 
    curve in the upper panel shows the combined result. The lower panel shows
    the improvement factor over the inner pulsar. \reply{The predicted positions of the degeneracy peaks from
    Eq.~(\ref{eq:chi}) are denoted with vertical lines.}}
    \label{fig:inc}
\end{figure}

As the second pulsar has a larger orbital period than the inner one, the 
constraint of the spin from it is in general looser than that from the inner
pulsar~\cite{Psaltis:2015uza,Zhang:2017qbb,Bower:2018mta,Hu:2023ubk}.  Thus the
improvement of the spin measurement by combining two pulsars is mainly caused by
the breaking of the leading-order degeneracy.  Figure~\ref{fig:inc} shows an
example of the dependence of the degeneracy on the most relevant orbital
parameter, the orbital inclination $i$. In this example we set $P_{b1}=2.3\,{\rm
yr}$, $e_1=0.8$, $P_{b2}=4.2\,{\rm yr}$, and $e_2=0.6\,$. The orbital
inclination $i_1$ of the inner pulsar is fixed to be $60^\circ$ and we vary the
orbital inclination of the second pulsar, $i_2$.  One can clearly see that the
improvement factor varies over two orders of magnitude and it is not dominated
by  the second pulsar's measurement precision. 

\reply{The simulation results are consistent with the prediction from the analysis of Eq.~(\ref{eq:chi}) as shown below.}  From Eq.~(\ref{eq:chi}), it is obvious that the shape of the
hyperbola of each pulsar is controlled by their inclination angles, which
determine the leading-order degeneracy. Moreover, from Eq.~(\ref{eq:chi}) one
can also predict the position of the three degeneracy peaks for the combined
results in Fig.~\ref{fig:inc}, where the improvement from combining a second
pulsar is the smallest. The central peak at $i=90^\circ$ originates from the 
denominator of the first term in Eq.~(\ref{eq:chi}), while the other two peaks 
are attributed to the situation where two hyperbolas are tangent to each other
at the point they intersect. We show the predicted peak positions from
Eq.~(\ref{eq:chi}) in Fig.~\ref{fig:inc} with vertical lines. 
\reply{Note that $s=0$ and $s=1/\sqrt{3}$ do not lead to additional degeneracy peaks as they behave normally when considering error propagation in Eq.~(\ref{eq:chi}).}

\textit{Pulsars in the GC.}---We have shown the large potential of measuring the
spin of Sgr~A* with two or more pulsars. Compared to the measurement with a
single pulsar, the key advantage of the approach we proposed here is that it
does not require a pulsar in an extremely compact orbit around Sgr~A*. Although
several pulsars including a magnetar and a millisecond pulsar have been found in
the GC region~\cite{Eatough:2013nva, Rea:2013pqa, Lower:2024sdi,
Johnston:2006fx, Deneva:2009mx, Abbate:2023car}, none of them are close enough
to the Sgr~A* for doing the measurement described above. Here we give an
estimation of the probability of finding suitable pulsars in future
high-frequency radio surveys.

We define two possibilities $P_1$ and $P_2$ as follows. $P_1$ represents the 
probability of finding the innermost pulsar with an orbital period $P_b\leq
0.5\,{\rm yr}$  via future instruments like SKA, which then can be used to
measure the spin of Sgr~A* by timing itself alone, while $P_2$ represents the
probability of finding two innermost  pulsars both with orbital period $P_b\leq
5\,{\rm yr}$. By combining the timing observations of these two pulsars, one may
constraint the spin of Sgr~A* to $\sim 1\%$ within five years according to our
simulations.  We assume a simple power-law distribution of pulsars near the
Sgr~A*, $p(r) \propto r^{-\alpha}$, where $p(r)$ is the probability density of
finding a pulsar in a unit volume at a distance $r$ from the Sgr~A*. 
Considering the large uncertainties in the GC pulsar population, the power-law
index $\alpha$ can vary largely depending on the detailed
models~\cite{Zhang:2014kva}. 
Under these assumptions, one can calculate $P_1$ and $P_2$ via
\begin{eqnarray}
    P_1&=&1-\exp\left[-N_4\left(\frac{r_{0.5\,{\rm yr}}}{R_4}\right)^{3-\alpha}\right]
    \,,\label{eq:P1}\\
    P_2&=&1-\left[1+N_4\left(\frac{r_{5\,{\rm yr}}}{R_4}\right)^{3-\alpha}\right]\exp
    \left[-N_4\left(\frac{r_{5\,{\rm yr}}}{R_4}\right)^{3-\alpha}\right]\,,
    \label{eq:P2}
\end{eqnarray}
where $r_{0.5\,{\rm yr}}\sim 100\,{\rm AU}$ and $r_{5\,{\rm yr}}\sim 475\, {\rm
AU}$ are the semimajor axes of the pulsar orbits with $P_b = 0.5 \,{\rm yr}$ and
$P_b = 5\,{\rm yr}$ respectively. $N_4$ is the expected total number of pulsars
that can be observed within $R_4=4000\,{\rm AU}$ around Sgr~A*.  $P_1$ and $P_2$
for different $\alpha$ and $N_4$, motivated from current observations and
simulations, are listed in Table~\ref{tab:prob}~\cite{Zhang:2014kva,
EHT:2023hcj}. In the reasonable parameter space of $\alpha$ and $N_4$, one
always has $P_1\lesssim P_2$,  indicating that it is more likely to find two
pulsars that are suitable for measuring the spin of Sgr A*.

\begin{table}
    \begin{center}
        \caption{$P_1$ and $P_2$ calculated with Eq.~(\ref{eq:P1}) and
        Eq.~(\ref{eq:P2}) for different $\alpha$ and $N_4$. In the reasonable
        parameter space, it is more likely to find two pulsars with orbital
        periods $P_b\leq 5\,{\rm yr}$ rather than a pulsar in a tight orbit that
        $P_b\leq 0.5\,{\rm yr}$.}\label{tab:prob}
        \renewcommand\arraystretch{1.2}
        \begin{tabularx}{\linewidth}{X<{\centering}X<{\centering}X<{\centering}
            X<{\centering}X<{\centering}X<{\centering}X<{\centering}
            X<{\centering}X<{\centering}X<{\centering}}
            \hline\hline
            $\alpha$ & $N_4$ & $P_1$ & $P_2$ & $N_4$ & $P_1$ & $P_2$ & $N_4$ 
            & $P_1$ & $P_2$\\
            \hline
            $1.6$ & $5$ & $2.9\%$ & $2.7\%$ & $10$ & $5.7\%$ & $9.2\%$ & $20$ & 
            $11\%$ & $27\%$\\
            $2.0$ & $5$ & $12\%$  & $12\%$  & $10$ & $23\%$  & $33\%$  & $20$ & 
            $40\%$ & $69\%$\\
            $2.4$ & $5$ & $43\%$  & $41\%$  & $10$ & $67\%$  & $77\%$  & $20$ & 
            $89\%$ & $98\%$\\
            \hline
        \end{tabularx}
    \end{center}
\end{table}

\textit{Discussions.}---In this {\it Letter}, we proposed to measure the spin of
Sgr~A* with timing observations of two or more pulsars, which does not require a
pulsar with an extremely small orbital period but can still provide similar
measurement precision by breaking the leading-order degeneracy of spin
parameters. Although the GC pulsar population has a large uncertainty at
present, it is expected to find a lot of new pulsars in future high-frequency
radio surveys towards the GC and we give a rough estimation of the probability
of finding proper systems.  It seems that we are more likely to find the systems
that are suitable for the procedure proposed here. Even if we can find a pulsar
with an orbital period $P_b\lesssim 0.5\,{\rm yr}$, combining a proper second
pulsar can still provide an improvement factor  $\sim 2$ for the SMBH spin
measurement.

In this work we have ignored the complex environments near the Sgr~A*, which
will complicate the measurement~\cite{Merritt:2009ex, Liu:2011ae,
Psaltis:2015uza, Hu:2023ubk}. In general, pulsars with larger orbital periods
will suffer from more perturbations caused by the mass distributions around
Sgr~A*, which may even spoil the measurement. Estimations made by previous studies suggest that for pulsars with orbital period $P_b\lesssim 3\,{\rm yr}$, the orbital advance rate caused by the SMBH spin is still larger than that caused by the stellar perturbation, which indicates the measurability of the SMBH spin with pulsar-SMBH systems~\cite{Liu:2011ae}. It is also suggested that one may only
use the timing data around the periastron passages and treat each periastron
passage incoherently~\cite{Psaltis:2015uza}. This will  enlarge the measurement
uncertainty of the spin from timing a single pulsar, where second-order time
derivatives are required but hard to  measure during the short periastron
passage. However, as only the leading-order time derivatives are required in the
method we proposed here, it is expected to have a good measurement even only the
timing data around the periastron passages are used. On the other hand, by
combing the timing observations of more pulsars, one will have the ability to
model the mass distribution around Sgr~A* that will be of vastly useful for
studies of the GC region.

\begin{acknowledgments}
We thank Norbert Wex, Ziri Younsi, and Fupeng Zhang for helpful discussions, \reply{and the anonymous referee for helpful comments.}
This work was supported by the National SKA Program of China (2020SKA0120300),
the National Natural Science Foundation of China (11991053), the Beijing Natural
Science Foundation (1242018), the Max Planck Partner Group Program funded by the
Max Planck Society, and the High-performance Computing Platform of Peking
University. 
\end{acknowledgments}

\bibliography{refs}

\end{document}